\documentclass[conference,letterpaper]{IEEEtran}
\IEEEoverridecommandlockouts

\def\Figs{./figs/} 

\usepackage{graphicx}
\usepackage{balance}
\usepackage[utf8]{inputenc} 
\usepackage[T1]{fontenc}
\usepackage{url}
\usepackage{ifthen}
\usepackage{comment}
\usepackage[cmex10]{amsmath} 

\newboolean{arXiv} 
\setboolean{arXiv}{false} 

\newboolean{NOTE} 
\setboolean{NOTE}{true} 

\usepackage[url,hyperrefblack,notheorems,IEEEtran]{research17} 
\usepackage{amsmath,amssymb,amsfonts,amsbsy}
\usepackage{amsthm} 
\newtheorem{theorem}{\mytheoremname}

\newtheorem{definition}{\mydefinitionname}

\usepackage{subcaption} 
\usepackage[capitalize]{cleveref}
\allowdisplaybreaks

\usepackage{array}
\usepackage{multirow}
\usepackage{enumitem} 
\usepackage{nicematrix} 
\usepackage{pgfplots}
\pgfplotsset{compat=newest}
\usepackage{tikz}
\usetikzlibrary{positioning}
\newcommand{\bigOop}{\operatorname{\mathcal{O}}} 

\newcommand{\BigbigOf}[1]{\bigOop\Bigl(#1\Bigr)}

\newcommand{\invQop}{\inv{\operatorname{\mathcal{Q}}}} 

\newcommand{\eQfinv}[1]{\invQop(#1)}

\newcommand{\dTV}[2]{d_{\textnormal{TV}}\left(#1 \kern0.1em{,}\kern0.1em #2\right)}
\newcommand{\edTV}[2]{d_{\textnormal{TV}}(#1 \kern0.1em{,}\kern0.1em #2)}
\newcommand{\bigdTV}[2]{d_{\textnormal{TV}}\bigl(#1 \kern0.1em{,}\kern0.1em #2\bigr)}
\newcommand{\BigdTV}[2]{d_{\textnormal{TV}}\Bigl(#1 \kern0.1em{,}\kern0.1em #2\Bigr)}
\newcommand{\biggdTV}[2]{d_{\textnormal{TV}}\biggl(#1 \kern0.1em{,}\kern0.1em #2\biggr)}
\newcommand{\BiggdTV}[2]{d_{\textnormal{TV}}\Biggl(#1 \kern0.1em{,}\kern0.1em #2\Biggr)}

\newcommand*{\Scale}[2][4]{\scalebox{#1}{\ensuremath{#2}}} 
\usepackage{soul} 
\usepackage[colorinlistoftodos, textsize=footnotesize]{todonotes} 
\newcommand{\ys}{\color{magenta}} 
\begin{document}
\title{Achieving Optimal Short-Blocklength\\ Secrecy Rate Using Multi-Kernel PAC Codes\\ for the Binary Erasure Wiretap Channel}


\author{%
  \IEEEauthorblockN{Hsuan-Yin Lin}
  \IEEEauthorblockA{Simula UiB, N--5006 Bergen, Norway
    \\
    Email: lin@simula.no}
  \and
  \IEEEauthorblockN{Yi-Sheng Su and Mao-Ching Chiu}
  \IEEEauthorblockA{Department of Communications Engineering
    \\ 
    National Chung Cheng University, Taiwan
    \\
    Email: \{yishengsu, ieemcc\}@ccu.edu.tw}
}


\maketitle


\begin{abstract}
    We investigate practical short-blocklength coding for the semi-deterministic binary erasure wiretap channel (BE-WTC), where the main channel to the legitimate receiver is noiseless, and the eavesdropper's channel is a binary erasure channel (BEC). 
    It is shown that under the average total variation distance secrecy metric, \emph{multi-kernel polarization-adjusted convolutional (MK-PAC)} codes can achieve the best possible theoretical secrecy rate at blocklengths of 16, 32, 64, and 128 if the secrecy leakage is less than or equal to certain values.
\end{abstract}

\section{Introduction}
\label{sec:introduction}

In the seminal paper by Wyner in 1975~\cite{Wyner75_1}, it was established that in the presence of an eavesdropper, keyless confidential and reliable communication between two legitimate parties is possible at rates up to the so-called \emph{secrecy capacity} of a wiretap channel (WTC). Since then, secrecy capacities of general WTCs have been characterized~\cite{BlochBarros11_1, OggierHassibi11_1}. 
Correspondingly, numerous coding proposals have appeared for very large blocklengths. Secrecy capacity-achieving coding schemes have been developed by employing low-density parity-check (LDPC) codes~\cite{ThangarajDihidarCalderbankMcLaughlinMerolla07_1, SubramanianThangarajBlochMcLaughlin11_1, RathiAnderssonThobabenKliewerSkoglund13_1}, polar codes~\cite{AnderssonRathiThobabenKliewerSkoglund10_1, MahdavifarVardy11_1}, and lattice codes~\cite{LingLuzziBelfioreStehle14_1}. Especially, LDPC codes were shown to achieve very good secrecy performance at large blocklengths in terms of the (normalized) \emph{equivocation} measure for the binary erasure WTCs~\cite{SubramanianThangarajBlochMcLaughlin11_1, RathiAnderssonThobabenKliewerSkoglund13_1}.

However, in modern emerging communication systems, e.g., smart-traffic safety and machine-to-machine communication, non-asymptotic secrecy rates are of paramount importance, as conventional coding schemes designed for large blocklengths result in long latency delays. In this respect, the recent contribution~\cite{YangSchaeferPoor19_1} is of notable interest, where non-asymptotic information theoretic rates were derived, accounting jointly for reliability and secrecy constraints at finite blocklengths. To date, there are, however, only a handful of works on coding for secrecy in the finite blocklength regime~\cite{PfisterGomesVilelaHarrison17_1, HarrisonBloch19_1, Shakiba-HerfehLuzziChorti21_1, OggierSoleBelfiore16_1, BollaufLinYtrehus23_3, BollaufLinYtrehus24_1}.

Recently, by concatenating an outer rate-$1$ convolutional code with the polar transform~\cite{Arikan09_1}, polarization-adjusted convolutional (PAC) code was proposed in~\cite{Arikan19_1sub} and has been shown significant advantages for the classical one-to-one noisy channel. It has been shown that PAC codes can almost achieve the \emph{normal approximation} bound~\cite{PolyanskiyPoorVerdu10_1} at short blocklengths~\cite{Arikan19_1sub, ChiuSu23_1}. Motivated by the remarkably good performance of PAC codes at short blocklengths, we use PAC codes based on mixing multiple kernels of different sizes, termed \emph{multi-kernel PAC (MK-PAC) codes}, to design a wiretap coding scheme.

In this work, we consider the average \emph{total variation distance (TVD)} as the secrecy metric~\cite{YangSchaeferPoor19_1} and derive the non-asymptotic theoretical bounds on the secrecy rate (Theorem~\ref{thm:BE-WTC-nonasymptotics}) for the semi-deterministic binary erasure wiretap channel (BE-WTC), where the main channel is noiseless, and the eavesdropper’s channel is a BEC. We further provide the achievable secrecy rates of MK-PAC codes and compare them to the second-order secrecy rates, the random coding achievability, and the exact converse bounds. The results show that under the average TVD secrecy metric, MK-PAC codes can achieve secrecy rates beyond the second-order approximation rate for short blocklengths. In particular, we observe that MK-PAC codes can achieve the optimal secrecy rate, i.e., the converse bound for secrecy rate, at blocklengths $n=16, 32, 64$, and $128$ when the secrecy leakage does not exceed $0.001$ (see Fig.~\ref{fig:secrecy-rates_delta1e-3}). Moreover, we present additional evidence with secrecy leakage bounded from above by $0.01$ to support the above observation, indicating that MK-PAC codes can also achieve the optimal secrecy rate for $n=16,32$, and $64$ (see Fig.~\ref{fig:secrecy-rates_delta1e-2}). To the best of our knowledge, this is the first work that demonstrates the optimal secrecy performance in short blocklengths.

\section{Preliminaries and Channel Model}
\label{sec:preliminaries-channel-model}

\subsection{Notation}
\label{sec:notation}

We denote by $\Naturals$ the set of all positive integers, and $[a:b]\eqdef\{a,a+1,\ldots,b\}$ for $a,b\in\{0\}\cup\Naturals$, $a \leq b$. Unless otherwise specified in the context, row-wise vectors are denoted by bold letters, matrices by sans-serif letters, random variables (RVs) (either scalar or vector) by capital letters, and sets by calligraphic capital letters, e.g., $\vect{x}$, $\mat{X}$, $X$, and $\set{X}$, respectively. 
The all-one (all-zero) row vector is denoted by $\vect{1}$ ($\vect{0}$), and its length will be clear from the context. When a set of indices $\set{S}$ is given, $\vect{x}_{\set{S}}$ denotes $\{\vect{x}_s\colon s\in\set{S}\}$. $\E[X]{\cdot}$ denotes expectation with respect to the RV $X$. $X\sim P_X$ denotes an RV distributed according to a probability mass function (PMF) $P_X(x)$, $x\in\set{X}$, and $\mathrm{U}_{\set{S}}$ represents a uniform distribution over a set $\set{S}$. $\HP{\cdot}$ denotes the entropy function, $\trans{(\cdot)}$ the transpose of a matrix.

\subsection{Wiretap Coding, Polar, Reed--Muller, and PAC Codes }
\label{sec:wiretap-coding_Polar-ReedMuller-PAC-codes}

We first introduce the notion of wiretap coding.
\begin{definition}[Wiretap Codes~\cite{BlochBarros11_1,YangSchaeferPoor19_1}]
  An $(n,\const{M},\eps,\delta)$ wiretap coding scheme for a discrete memoryless wiretap channel (DM-WTC) $(\set{X},\set{Y}\times\set{Z},P_{Y,Z\mid X})$ consists of
  \begin{itemize}
  \item a message $M$, which is assumed to be uniformly distributed on the message set $\set{M}\eqdef[1:\const{M}]$,
  \item an encoding function $f\colon\set{M}\rightarrow\set{X}^n$ that maps each message $m\in\set{M}$ into the corresponding
    length-$n$ codeword $\vect{x}_m\in\set{X}^n$, $n\in\Naturals$,
  \item a decoding function $g\colon\set{Y}^n\to [1:\const{M}]$ that makes a decoding decision $g(\vect{y})=\hat{m}\in\set{M}$ for every received $n$-vector $\vect{y}\in\set{Y}^n$,
  \end{itemize}
  and the code should satisfy the average error probability constraint
  \begin{equation}
    \Prv{g(\vect{Y})\neq M}\leq\eps,
    \label{eq:constraint_average-error}
  \end{equation}
  and the average TVD secrecy metric constraint
  \begin{equation}
    \bigdTV{P_{M,Z}}{\mathrm{U}_{\set{M}} P_Z}\leq\delta,
    \label{eq:constraint_secrecy_rate}
  \end{equation}
  where $\bigdTV{P}{Q}\eqdef\frac{1}{2}\sum_{x\in\set{X}}\abs{P(x)-Q(x)}$.
\end{definition}

\begin{definition}[Maximal Secrecy Rate]
  \label{def:def_maximal-secrecy-rate}
  The largest possible secrecy rate under average error probability and average TVD constraints is defined as
  \begin{equation*}
    \const{R}^\ast(n,\eps,\delta)\eqdef\max\biggl\{\frac{\log{\const{M}}}{n}\colon\exists\, (n,\const{M},\eps,\delta) \textnormal{ wiretap code}\biggr\}.
  \end{equation*}
\end{definition}

In general, potential candidates for practical wiretap code constructions include LDPC codes, polar codes, and lattice codes. In this work, the finite-blocklength secrecy-good wiretap codes are developed based on polar codes and their extensions: the PAC codes.

\subsubsection*{Polar, Reed-Muller, and PAC Codes}
\label{sec:Polar-PAC-ReedMuller-codes}

Polar codes were invented by Arikan and proved to achieve the capacity of arbitrary symmetric discrete memoryless channels (DMCs) with the low-complexity successive cancellation decoder \cite{Arikan09_1}. The generator matrix of polar codes with blocklength $n=2^s$ is defined as
\begin{equation*}
  \mat{G}_{\textnormal{polar}} = \mat{G}_2^{\otimes s}, 
\end{equation*}
where $\mat{G}_2 = \bigl[\begin{smallmatrix} 1 & 0 \\ 1 & 1 \end{smallmatrix}\bigr]$ and $\otimes s$ denotes the $s$-th Kronecker power. Polar codes transfer the original $n$ identical independent copies of a DMC into $n$ synthesized channels. On the other hand, Reed--Muller codes~\cite[Ch.~13]{MacWilliamsSloane77_1} 
were shown to achieve the capacity of the BEC in~\cite{Kudekar-etal17_1}. Reed--Muller codes have the same encoding structure as polar codes. The difference between the Reed--Muller and polar codes is the selection rule of synthesized channels. Another powerful family of codes utilizing the polarization effect is the PAC codes. It has been shown that the design of PAC codes based on the Reed--Muller rule achieves remarkable performance at short blocklengths. A PAC code is the concatenation of a rate-1 outer convolutional code and an inner polar code. Let variable $\const{D}$ represent a unit time delay. The generator polynomial of the convolutional code can be represented as $p(\const{D}) = p_0+p_1\const{D} + \ldots + p_{\nu} \const{D}^{\nu}$ with $p_0=p_{\nu} = 1$. The parameter $\nu+1$ is termed the constraint length of the convolutional code. Let $\mat{P}$ be the $n \times n$ upper-triangular generator matrix of the convolutional code, i.e.,
\begin{equation*}
  \mat{P} =
  \begin{bmatrix}
    \begin{array}{cccccccc}
      1 & p_1 & \ldots & p_{\nu-1} & 1 & 0 & \cdots & 0 \\
      0 & 1   & p_1 & \cdots & p_{\nu-1} & 1 & \cdots & 0 \\
      \vdots & \vdots & \ddots & \ddots & \ddots & \ddots & \vdots &\vdots\\
      \vdots & \vdots& \vdots& \vdots & \ddots & \ddots &\vdots &\vdots \\
      0 & 0 & 0 & 0 & 0 & \cdots & \cdots & 1 \\
    \end{array}
  \end{bmatrix}.
\end{equation*}
Moreover, the generator matrix of PAC codes with blocklength $n=2^s$ is defined as
\begin{equation*}
  \mat{G}_{\textnormal{PAC}} = \mat{P}\cdot\mat{G}_{\textnormal{polar}}.
\end{equation*}

\subsection{The Semi-Deterministic Binary Erasure WTC (BE-WTC)}
\label{sec:BE-WTC}

\begin{figure}[t!]
  \Scale[0.95]{
    \includegraphics{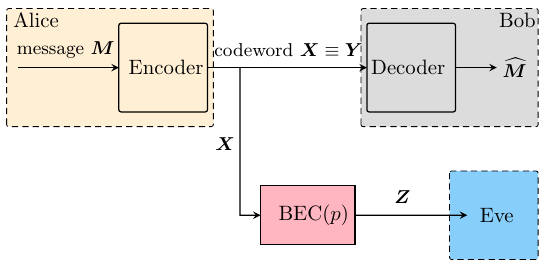}
  }
  \caption{A semi-deterministic binary erasure WTC (BE-WTC).}
  \label{fig:fig_BE-WTC}
\end{figure}

This work mainly focuses on a simple but insightful DM-WTC, the semi-deterministic binary erasure WTC (BE-WTC) $P_{Y, Z\mid X}\colon\set{X}=\{0,1\}\rightarrow\set{Y}\times\set{Z}=\{0,1\}\times\{0,1,2\}$. The channel model is depicted in Fig.~\ref{fig:fig_BE-WTC}, where the main channel between $X$ and $Y$ is a noiseless channel and the eavesdropper's channel $P_{Z\mid X}$ is a BEC with erasure probability $0\leq p<1$, and the conditional channel law
\begin{IEEEeqnarray*}{c}
  P_{Z\mid X}(z|x)=
  \begin{cases}
    1-p & \textnormal{if } z=x;
    \\
    p & \textnormal{if } z=2,
  \end{cases}
  \quad x\in\{0,1\}.      
\end{IEEEeqnarray*}

Note that since the main channel is noiseless for the semi-deterministic DM-WTC, the average error probability in~\eqref{eq:constraint_average-error} is zero. Hence, the $\eps$ in the notation $\const{R}^\ast(n,\eps,\delta)$ and $(n,\const{M},\eps,\delta)$ can be omitted.

It is known that the secrecy capacity of the BE-WTC is
\begin{IEEEeqnarray*}{c}
  \const{C}_{\textnormal{BE-WTC}}=1-(1-p)=p,
\end{IEEEeqnarray*}
and the non-asymptotic second-order secrecy rate is (see \cite[eq.~(139)]{YangSchaeferPoor19_1})
\begin{IEEEeqnarray*}{c}
  \const{R}^\ast(n,\delta)=\const{C}_\textnormal{BE-WTC}-\sqrt{\frac{p(1-p)}{n}}\eQfinv{\delta}+\BigbigOf{\frac{\log{n}}{n}},
\end{IEEEeqnarray*}
where $\eQfinv{\cdot}$ is the inverse of the $\Q$-function $\Qop(\alpha)\eqdef\frac{1}{\sqrt{2\pi}}\int_{\alpha}^\infty\exp{\bigl(-\frac{t^2}{2}\bigr)}\dd t$.

\section{Main Results}
\label{sec:main-results}

\subsection{Non-Asymptotic Fundamental Limits on Secrecy Rate}
\label{sec:non-asymptotic-fundamental-limits_secrecy-rate}

We present the following non-asymptotic theoretical results on the secrecy performance for the BE-WTC, which can be proved using a proof similar to that for the semi-deterministic binary symmetric WTC~\cite[Th.~18]{YangSchaeferPoor19_1}.
\begin{theorem}
  \label{thm:BE-WTC-nonasymptotics}
  Consider a semi-deterministic BE-WTC with erasure probability $0\leq p< 1$. There exists a binary $(n,2^k,\delta)$ wiretap code such that
  \begin{IEEEeqnarray}{c}
    \delta\leq\frac{1}{2}\min_{\gamma>0}\Biggl\{g_n(\gamma)+\sqrt{g_n^2(\gamma)+\frac{\gamma}{2^{n-k}}h_n(\gamma)}\Biggr\},
    \label{eq:achievability_BEWTC}
  \end{IEEEeqnarray}
  where $k\in\Naturals$, and
  \begin{IEEEeqnarray*}{rCl}
    g_n(\gamma)& \eqdef &1-\bigE{2^{-\max\{n-B(n,p)-\log_2{\gamma},0\}}},
    \\[1mm]
    h_n(\gamma)& \eqdef &\bigE{2^{-\abs{n-B(n,p)-\log_2{\gamma}}}}.
  \end{IEEEeqnarray*}
  Here, $B(n,p)$ is the binomial RV with parameters $n$ and $p$. Conversely, every binary $(n,\const{M},\delta)$ wiretap code must satisfy
  \begin{IEEEeqnarray}{c}
    g_n\biggl(\frac{2^n}{\const{M}}\biggr)\leq\delta.
    \label{eq:converse-bound_BEWTC}
  \end{IEEEeqnarray}
\end{theorem}

\subsection{Wiretap Coding Scheme}
\label{subsec:wiretap-coding-scheme}
In this paper, we consider a wiretap coding scheme based on MK-PAC codes.

We first introduce the \emph{multi-kernel polar (MK-polar)} codes as follows~\cite{BioglioGabryLandBelfiore20_1}. MK-polar codes are a generalization of Arikan's polar codes, which are obtained by using binary kernels of different sizes to construct the generator matrix of the code. The generator matrix of MK-polar codes with blocklength $n$ is defined as
\begin{equation*}
  \mat{G}_{\textnormal{MK-polar}} = \mat{G}_{k_1} \otimes \mat{G}_{k_2} \otimes \cdots \otimes \mat{G}_{k_s},
\end{equation*}
where $n = k_1 k_2 \cdots k_s$ for some $s\in\Naturals$ and $\otimes$ denotes the Kronecker product. The generator matrix $\mat{G}_{\textnormal{MK-polar}}$ is the Kronecker product of $s$ polarization matrices $\mat{G}_{k_i}$ of size $k_i \times k_i$ with binary entries, called kernels of dimension $k_i$, $i\in [1:s]$. Throughout the paper, we will refer to polar codes when the Kronecker product of $\mat{G}_{\textnormal{MK-polar}}$ comprises only kernels $\mat{G}_2$, according to the original formulation by Arikan, while we will refer to MK-polar codes if $\mat{G}_{\textnormal{MK-polar}}$ comprises more than one kind of kernel.

An MK-PAC code is a PAC code with an MK-polar code as the inner code. The generator matrix of an MK-PAC code with blocklength $n$ is defined as
\begin{equation*}
  \mat{G}_{\textnormal{MK-PAC}} = \mat{P} \cdot \mat{G}_{\textnormal{MK-polar}}.
\end{equation*}
Consider a binary-input channel $\mat{W}\colon \set{X} = \left\{ 0, 1 \right\} \rightarrow \set{Z}$ and the corresponding channel with $n$ independent channel uses of $\mat{W}$, denoted by $\mat{W}^n$. Let the encoded codeword $\vect{x} = \vect{u} \mat{G}_{\textnormal{MK-PAC}}$ be the channel input and $\vect{z}$ be the channel output. The so-called \emph{bit-channel} $\mat{W}^{(i)}\equiv\mat{W}^{(i)}(\vect{z}, \vect{u}_{[1:i-1]}\mid u_i)$ which takes a single bit $u_i$ as input and the observation vector $\vect{z}$ and the past inputs $\vect{u}_{[1:i-1]}$ of $\mat{W}^n$ as output, is defined as
\begin{IEEEeqnarray*}{rCl}
  \IEEEeqnarraymulticol{3}{l}{%
    \mat{W}^{(i)}(\vect{z},\vect{u}_{[1:i-1]}\mid u_i)}\nonumber\\*\quad%
  & \eqdef &\frac{1}{2^{n-1}}\sum_{\vect{u}_{[i+1:n]}\in\{0,1\}^{n-i}}\widetilde{\mat{W}}(\vect{z}\mid \vect{u}_{[1:i-1]},u_i,\vect{u}_{[i+1:n]}),
\end{IEEEeqnarray*}
where $\widetilde{\mat{W}}(\vect{z}\mid\vect{u})\eqdef\mat{W}^n(\vect{z}\mid\vect{u}\mat{G}_{\textnormal{MK-PAC}})$, $i\in [1:n]$. It is shown that when the blocklength $n$ increases to infinity, the bit-channels $\mat{W}^{(i)}$ for polar codes, \emph{polarize}~\cite{Arikan09_1}, i.e., they are either noiseless or completely noisy. Given a blocklength $n$, we call those bit-channels almost noiseless \emph{good} bit-channels, and those that are almost completely noisy the \emph{poor} bit-channels.

We use a similar wiretap coding approach as~\cite[Secs.~III and IV]{MahdavifarVardy11_1}, where polar codes have been shown to asymptotically achieve the secrecy capacity for a large family of WTCs. Note that for the semi-deterministic WTC model, we don't need to consider the main channel, and only the eavesdropper's WTC $\mat{W}$ needs to be considered to build the index sets based on \cite[Sec.~IV]{MahdavifarVardy11_1}. Let a message $m\in\set{M}$ represented by a length-$k$ vector $\vect{m}$. The general idea is to transmit the message $\vect{m}$ \emph{only} through those poor bit-channels to the eavesdropper. More specifically, let us define the \emph{Bhattacharyya parameter} of the $i$-th bit channel $\mat{W}^{(i)}$ to be
\begin{IEEEeqnarray*}{c}
  \const{Z}(\mat{W}^{(i)})\eqdef\sum\limits_{\substack{\vect{z}, \\ \vect{u}_{[1:i-1]}}} 
  \sqrt{\mat{W}^{(i)}(\vect{z},\vect{u}_{[1:i-1]}\mid 0)\mat{W}^{(i)}(\vect{z},\vect{u}_{[1:i-1]}\mid 1)}.\IEEEeqnarraynumspace
\end{IEEEeqnarray*}
We then construct a specific index subset $\set{A}\subset\{1, 2,\ldots,n\}$ with cardinality $\ecard{\set{A}} = k$, such that for all $i \in \set{A}$ and $j \in \cset{\set{A}}$, we have $\const{Z}(\mat{W}^{(i)})\geq \const{Z}(\mat{W}^{(j)})$. Next, the input $\vect{U}$ of the encoder is assigned to be $\vect{U}_{\set{A}} = \vect{m}$ and $\vect{U}_{\cset{\set{A}}} = \vect{V}$, where $\vect{V}$ is a random vector of $n-k$ independent and identically distributed uniform binary RVs. Consequently, the transmitted codeword is $\vect{X} = \vect{U}\mat{G}_{\textnormal{MK-PAC}}=\vect{U}\mat{P}\mat{G}_{\textnormal{MK-polar}}$.

The above wiretap coding scheme can be regarded as a special case of \emph{coset coding}. To see this, given an index set $\set{A}$ with $\ecard{\set{A}}=k$ and a fixed vector $\vect{m}\in\{0,1\}^k$ of length $k$, we define $\code{C}(\vect{m},\set{A})$ as the binary linear code such that $\code{C}(\vect{m},\set{A})\eqdef\{\vect{x}=\vect{u}\mat{G}_{\textnormal{MK-PAC}}: \vect{u}_{\set{A}}=\vect{m}, \vect{u}_{\cset{\set{A}}}\in\{0,1\}^{n-k}\}$. It follows that $\code{C}(\vect{m},\set{A})$ is a coset code of $\code{C}(\vect{0},\set{A})$. Hence, we have
\begin{equation*}
  \code{C} = \bigcup_{\vect{m}\in \{0,1\}^k} \code{C}(\vect{m},\set{A}) = \{0,1\}^n,
\end{equation*}
and $\code{C}(\vect{m},\set{A})\subseteq\code{C}$ forms a \emph{nested} code structure~\cite[Ch.~6]{BlochBarros11_1}.

\ifthenelse{\boolean{arXiv}}{
{\ys
\subsection{Average TVD Secrecy Metric for the BE-WTC}
\label{subsec:average-TVD-BE-WTC}

\begin{theorem}
  \label{thm:BE-WTC-avgTVD}
  Consider a semi-deterministic BE-WTC with erasure probability $0\leq p< 1$. Suppose that $\code{C}_0$ is a maximum distance separable (MDS) code. Then, the average TVD secrecy metric has the following closed-form expression:
  \begin{align}
    \bigdTV{P_{M,\vect{Z}}}{\mathrm{U}_{\set{M}} P_{\vect{Z}}}= 
    \sum_{e=0}^{k-1} \left(1-\frac{1}{2^{k-e}} \right) {n \choose e} p^e (1-p)^{n-e}.
    \label{eq:avgTVD_BE_WTC}
  \end{align}
\end{theorem}
\begin{IEEEproof}
 Let $\vect{z} \in \set{Z}^n = \left\{ 0, 1, 2 \right\}^n$ be an eavesdropper's observation and $e$ the number of erased bits in $\vect{z}$. With a slight abuse of notation, let $\vect{x}_{\vect{z}}$ be a codeword that agrees with $\vect{z}$ in the $n-e$ unerased positions. Thus $\vect{x}_{\vect{z}}$ is said to be consistent with $\vect{z}$. If a coset of $\code{C}_0$ contains at least one codeword that is consistent with $\vect{z}$, the coset is said to be consistent with $\vect{z}$. Given message $m$ and observation $\vect{z}$, the joint distribution $P_{M,\vect{Z}}(m, \vect{z})$ can be determined as follows:
 \begin{eqnarray}
     \lefteqn{P_{M,\vect{Z}}(m, \vect{z})} \nonumber \\ 
     & = & P_{\vect{Z} \mid M}(\vect{z} \mid m) P_{M}(m) \nonumber \\
     & = & \sum_{\vect{x}_{\vect{z}} \in \set{X}^n} p^e (1-p)^{n-e} \frac{1}{2^{n-k}} \I{\vect{x}_{\vect{z}} \in \code{C}_{0}(\vect{m})} \cdot \frac{1}{2^k} \nonumber \\
     & = & \frac{p^e (1-p)^{n-e}}{2^n} \sum_{\vect{x}_{\vect{z}} \in \set{X}^n} \I{\vect{x}_{\vect{z}} \in \code{C}_{0}(\vect{m})}, 
     \label{eq:joint-dist}
 \end{eqnarray}
 where $\set{X} = \left\{ 0, 1 \right\}$ and $\I{\cdot}$ is the indicator function. To determine $P_{\vect{Z}}(\vect{z})$ and $\mathrm{U}_{\set{M}}(m) P_{\vect{Z}}(\vect{z})$, we consider the following two cases depending on the values of $e$:
 \begin{enumerate}
     \item{$e \in [0:k-1]$}: Since $\code{C}_0$ is an $(n, n-k)$ MDS codes, any $k$ columns of the generator matrix of $\code{C}_0$ are linearly independent. This indicates that, given $e=k$, the total number of cosets of $\code{C}_0$ consistent with $\vect{z}$ is exactly equal to $2^k$ and the number of codewords consistent with $\vect{z}$ in each consistent coset is equal to one. However, under the scenario $e \in [0:k-1]$, the total number of cosets of $\code{C}_0$ consistent with $\vect{z}$ is reduced to $2^e$. Thus, for $e \in [0:k-1]$, we have 
     \begin{align}
        P_{\vect{Z}}(\vect{z}) 
        = \sum_{m \in \set{M}} P_{M,\vect{Z}}(m, \vect{z}) 
        = \frac{p^e (1-p)^{n-e}}{2^{n-e}}
        \label{eq:z-marginal-dist}
     \end{align}
     and
     \begin{align}
        \mathrm{U}_{\set{M}}(m) P_{\vect{Z}}(\vect{z})  
        = \frac{1}{2^k} \cdot \frac{p^e (1-p)^{n-e}}{2^{n-e}}.
        \label{eq:product-marginal-dist}
     \end{align}
     Then, by (\ref{eq:joint-dist}) and (\ref{eq:product-marginal-dist}), we can determine
     \begin{align}
        & \sum_{m \in \set{M}}\sum_{\substack{\vect{z} \in \set{Z}^n \\ e \in [0:k-1]}}\abs{P_{M,\vect{Z}}(m, \vect{z})-\mathrm{U}_{\set{M}}(m) P_{\vect{Z}}(\vect{z})} \nonumber \\
        & = \sum_{e=0}^{k-1}\sum_{m \in \set{M}} \bigg| \frac{p^e (1-p)^{n-e}}{2^n} \sum_{\vect{x}_{\vect{z}} \in \set{X}^n}  \I{\vect{x}_{\vect{z}} \in \code{C}_{0}(\vect{m})} \nonumber \\
        & \hspace{25mm} - \frac{p^e (1-p)^{n-e}}{2^{n+k-e}} \bigg| \label{eq:abs-diff} \\
        & = \sum_{e=0}^{k-1} \left\{ {n \choose e} 2^e \left[ \frac{p^e (1-p)^{n-e}}{2^n} - \frac{p^e (1-p)^{n-e}}{2^{n+k-e}} \right] \right. \nonumber \\
        & \hspace{15mm} \left. + \left[ {n \choose e} 2^{n+k-e} - {n \choose e} 2^e \right] \frac{p^e (1-p)^{n-e}}{2^{n+k-e}} \right\} \nonumber \\
        & = \sum_{e=0}^{k-1} 2 \left( 1 - \frac{1}{2^{k-e}} \right) {n \choose e} p^e (1-p)^{n-e}.
        \label{eq:case1}
     \end{align}
     \item{$e \in [k:n]$}: Under this scenario, all cosets of $\code{C}_0$ are consistent with $\vect{z}$ and, in each consistent coset, the number of codewords consistent with $\vect{z}$ is equal to $2^{e-k}$. Thus, for $e \in [k:n]$, the total number of codewords consistent with $\vect{z}$ is equal to $2^k \cdot 2^{e-k} = 2^e$, so that (\ref{eq:z-marginal-dist}) and (\ref{eq:product-marginal-dist}) still hold. In addition, the summation $\sum_{\vect{x}_{\vect{z}} \in \set{X}^n} \I{\vect{x}_{\vect{z}} \in \code{C}_{0}(\vect{m})}$ in (\ref{eq:abs-diff}) must be equal to $2^{e-k}$. Thus, by (\ref{eq:abs-diff}), we have
     \begin{align}
         \sum_{m \in \set{M}}\sum_{\substack{\vect{z} \in \set{Z}^n \\ e \in [k:n]}}\abs{P_{M,\vect{Z}}(m, \vect{z})-\mathrm{U}_{\set{M}}(m) P_{\vect{Z}}(\vect{z})} = 0.
         \label{eq:case2}
     \end{align}
 \end{enumerate}
 Then, by combining (\ref{eq:case1}) and (\ref{eq:case2}), the average TVD secrecy metric is
 \begin{align*}
    & \bigdTV{P_{M,\vect{Z}}(m, \vect{z})}{\mathrm{U}_{\set{M}}(m) P_{\vect{Z}}(\vect{z})} \\
    & \hspace{5mm} = \frac{1}{2}\sum_{m \in \set{M}}\sum_{\vect{z}\in\set{Z}^n}\abs{P_{M,\vect{Z}}(m, \vect{z})-\mathrm{U}_{\set{M}}(m) P_{\vect{Z}}(\vect{z})} \\
    & \hspace{5mm} = \sum_{e=0}^{k-1} \left( 1 - \frac{1}{2^{k-e}} \right) {n \choose e} p^e (1-p)^{n-e}.
 \end{align*}
 This completes the proof.    
\end{IEEEproof}
}}{}

\section{Numerical Results}
\label{sec:numerical-results}

Here, the secrecy performance of MK-PAC codes for the BE-WTC at short and medium blocklengths are evaluated with numerical simulations.

\subsection{Average TVD of Bit-Channels for Multi-Kernel PAC Codes}
\label{sec:avgTVD}

In this subsection, the average TVD of each bit-channel $\mat{W}^{(i)}(\vect{z},\vect{u}_{[1:i-1]}\mid u_i)$ for the BE-WTC with erasure probability $p=0.4$ and blocklength $n = 128$ is presented. It is well-known that for $\mat{W}$ being a BEC, the bit-channels $\mat{W}^{(i)}$, $i \in [1:n]$, are also BECs~\cite[Fact 1]{Shakiba-HerfehLuzziChorti21_1}. It is also known from~\cite[Lemma 3]{Shakiba-HerfehLuzziChorti21_1} that the average TVD of a BEC with erasure probability $\tilde{p}$, uniform input $X$, and output $Y$, is $\bigdTV{P_{X, Y}}{\frac{1}{2} P_{Y}} = \frac{1}{2}(1-\tilde{p})$. Thus, the average TVD of each bit-channel $\mat{W}^{(i)}$ can be evaluated in terms of its erasure probability. The erasure probabilities of the bit-channels for MK-PAC codes are derived by using Monte-Carlo simulations with $2 \times 10^5$ channel realizations, under which the erasure probabilities are correct up to three decimal places.

We perform the Monte-Carlo simulations as follows. For each channel realization, we obtain a vector $\vect{z}'$ by omitting the erased bits of the observed vector $\vect{z}$ in that channel realization and a matrix $\mat{G}'_{\textnormal{MK-PAC}}$ by removing the columns of $\mat{G}_{\textnormal{MK-PAC}}$ that correspond to those erased bits. Since the past bits $\vect{u}_{[1:i-1]}$ are assumed to be given for decoding the bit-channel $\mat{W}^{(i)}$, they have no impact on the uncertainty of the bit-channel $\mat{W}^{(i)}$. Thus, the subvectors $\vect{u}_{[1:i-1]}$ and $\vect{u}_{[i:n]}$ of $\vect{u}$ correspond to the known and unknown bits, respectively. Further, denote by $\tilde{\mat{G}}'_{\textnormal{MK-PAC}}$ and $\bar{\mat{G}}'_{\textnormal{MK-PAC}}$ the corresponding submatrices of $\mat{G}'_{\textnormal{MK-PAC}}$ regarding $\vect{u}_{[1:i-1]}$ and $\vect{u}_{[i:n]}$, respectively. As a result,
\begin{equation}
  \vect{u}_{[i:n]}\bar{\mat{G}}'_{\textnormal{MK-PAC}} = \vect{z}' + \vect{u}_{[1:i-1]} \tilde{\mat{G}}'_{\textnormal{MK-PAC}},\label{eq:monte-carlo-simu}
\end{equation}
and the vector $\vect{z}' + \vect{u}_{[1:i-1]} \tilde{\mat{G}}'_{\textnormal{MK-PAC}}$ is known. The bit-channel $\mat{W}^{(i)}$ is noiseless if and only if one can solve $u_i$ from (\ref{eq:monte-carlo-simu}).

\begin{figure}[t!]
%
%
\definecolor{mycolor1}{rgb}{1.00000,0.00000,1.00000}%
\begin{tikzpicture}

\begin{axis}[%
width=0.80\columnwidth,
height=0.30\textheight,
at={(1.011in,0.642in)},
scale only axis,
xmin=0,
xmax=128,
xlabel style={font=\color{white!15!black}},
xlabel={$j$-th smallest},
ymode=log,
ymin=0.0001,
ymax=1,
yminorticks=true,
ylabel style={font=\color{white!15!black}},
ylabel={Sorted average TVD},
axis background/.style={fill=white},
xmajorgrids,
ymajorgrids,
yminorgrids,
legend style={at={(1.00,0.00)},
font = \scriptsize, 
anchor=south east, legend cell align=left, align=left,
draw=white!15!black
}
]
\addplot [color=red, mark=x, mark options={solid, red}]
  table[row sep=crcr]{%
1	0\\
2	6.328271e-15\\
3	1.265654e-14\\
4	5.062617e-14\\
5	7.578382e-13\\
6	7.105622e-11\\
7	2.374943e-08\\
8	1.591732e-07\\
9	3.182566e-07\\
10	6.361542e-07\\
11	1.231145e-06\\
12	2.45956e-06\\
13	7.125975e-06\\
14	9.209221e-06\\
15	1.192102e-05\\
16	2.375993e-05\\
17	8.464537e-05\\
18	0.0002179186\\
19	0.0004294735\\
20	0.0005427986\\
21	0.00112733\\
22	0.001356919\\
23	0.002215449\\
24	0.00376805\\
25	0.004282462\\
26	0.00602821\\
27	0.006869705\\
28	0.007093565\\
29	0.00803025\\
30	0.01292653\\
31	0.01778962\\
32	0.02313872\\
33	0.0264756\\
34	0.02887832\\
35	0.0324056\\
36	0.04892876\\
37	0.0507376\\
38	0.05447822\\
39	0.08176561\\
40	0.1011044\\
41	0.1037735\\
42	0.1120162\\
43	0.1172204\\
44	0.1186998\\
45	0.1528631\\
46	0.170835\\
47	0.1919832\\
48	0.2036357\\
49	0.2180534\\
50	0.2259728\\
51	0.231196\\
52	0.2638931\\
53	0.2659298\\
54	0.2756073\\
55	0.3226243\\
56	0.3246469\\
57	0.3361535\\
58	0.348572\\
59	0.3669705\\
60	0.3844759\\
61	0.39737\\
62	0.397899\\
63	0.400062\\
64	0.4117651\\
65	0.4423304\\
66	0.4462962\\
67	0.4472277\\
68	0.4487981\\
69	0.4633572\\
70	0.4655124\\
71	0.468097\\
72	0.4707305\\
73	0.4808629\\
74	0.4811408\\
75	0.4837898\\
76	0.4898719\\
77	0.491111\\
78	0.4924231\\
79	0.4941119\\
80	0.4941761\\
81	0.4947828\\
82	0.4957205\\
83	0.4959669\\
84	0.4969242\\
85	0.4984173\\
86	0.4985288\\
87	0.499244\\
88	0.4993839\\
89	0.4994742\\
90	0.4995586\\
91	0.499776\\
92	0.4998133\\
93	0.4998872\\
94	0.4999057\\
95	0.4999482\\
96	0.4999601\\
97	0.499974\\
98	0.4999863\\
99	0.4999869\\
100	0.4999918\\
101	0.4999931\\
102	0.4999953\\
103	0.4999966\\
104	0.4999983\\
105	0.4999987\\
106	0.4999994\\
107	0.4999997\\
108	0.5\\
109	0.5\\
110	0.5\\
111	0.5\\
112	0.5\\
113	0.5\\
114	0.5\\
115	0.5\\
116	0.5\\
117	0.5\\
118	0.5\\
119	0.5\\
120	0.5\\
121	0.5\\
122	0.5\\
123	0.5\\
124	0.5\\
125	0.5\\
126	0.5\\
127	0.5\\
128	0.5\\
};
\addlegendentry{Polar Codes}

\addplot [color=blue, mark=o, mark options={solid, blue}]
  table[row sep=crcr]{%
1	0\\
2	0\\
3	0\\
4	0\\
5	0\\
6	0\\
7	0\\
8	0\\
9	0\\
10	0\\
11	0\\
12	0\\
13	0\\
14	0\\
15	0\\
16	0\\
17	0\\
18	2.5e-06\\
19	2.5e-06\\
20	2.5e-06\\
21	2.5e-06\\
22	2.5e-06\\
23	5e-06\\
24	5e-06\\
25	5e-06\\
26	5e-06\\
27	5e-06\\
28	7.5e-06\\
29	2.25e-05\\
30	0.0011075\\
31	0.0023275\\
32	0.00391\\
33	0.0059425\\
34	0.006125\\
35	0.00949\\
36	0.009735\\
37	0.015385\\
38	0.023165\\
39	0.0239325\\
40	0.03554\\
41	0.0461925\\
42	0.0524325\\
43	0.0651075\\
44	0.0696725\\
45	0.0935275\\
46	0.0979775\\
47	0.129905\\
48	0.13076\\
49	0.1394725\\
50	0.178845\\
51	0.2196725\\
52	0.2480775\\
53	0.29034\\
54	0.3005625\\
55	0.3323325\\
56	0.343215\\
57	0.3791225\\
58	0.39967\\
59	0.4259075\\
60	0.447565\\
61	0.452745\\
62	0.4670125\\
63	0.478615\\
64	0.4870775\\
65	0.495365\\
66	0.4970725\\
67	0.49814\\
68	0.4988575\\
69	0.4991275\\
70	0.499445\\
71	0.4996675\\
72	0.4997625\\
73	0.4998\\
74	0.49985\\
75	0.499885\\
76	0.499895\\
77	0.499905\\
78	0.4999175\\
79	0.499925\\
80	0.4999375\\
81	0.4999425\\
82	0.4999475\\
83	0.499955\\
84	0.49996\\
85	0.4999625\\
86	0.499965\\
87	0.4999675\\
88	0.49997\\
89	0.4999775\\
90	0.4999825\\
91	0.4999825\\
92	0.49999\\
93	0.49999\\
94	0.4999925\\
95	0.499995\\
96	0.499995\\
97	0.499995\\
98	0.4999975\\
99	0.4999975\\
100	0.5\\
101	0.5\\
102	0.5\\
103	0.5\\
104	0.5\\
105	0.5\\
106	0.5\\
107	0.5\\
108	0.5\\
109	0.5\\
110	0.5\\
111	0.5\\
112	0.5\\
113	0.5\\
114	0.5\\
115	0.5\\
116	0.5\\
117	0.5\\
118	0.5\\
119	0.5\\
120	0.5\\
121	0.5\\
122	0.5\\
123	0.5\\
124	0.5\\
125	0.5\\
126	0.5\\
127	0.5\\
128	0.5\\
};
\addlegendentry{Reed-Muller Codes}

\addplot [color=mycolor1, mark=triangle, mark options={solid, mycolor1}]
  table[row sep=crcr]{%
1	0\\
2	0\\
3	0\\
4	0\\
5	0\\
6	0\\
7	0\\
8	0\\
9	0\\
10	0\\
11	0\\
12	0\\
13	0\\
14	0\\
15	2.5e-06\\
16	1.5e-05\\
17	1.75e-05\\
18	8.25e-05\\
19	0.0002075\\
20	0.0002325\\
21	0.00026\\
22	0.0003675\\
23	0.0004125\\
24	0.0006975\\
25	0.0007925\\
26	0.0011675\\
27	0.0015325\\
28	0.0027975\\
29	0.00298\\
30	0.00403\\
31	0.007655\\
32	0.01406\\
33	0.018425\\
34	0.02153\\
35	0.02937\\
36	0.03224\\
37	0.0380925\\
38	0.0507175\\
39	0.051335\\
40	0.0596525\\
41	0.0623325\\
42	0.06248\\
43	0.07375\\
44	0.07501\\
45	0.0811825\\
46	0.1139425\\
47	0.1539525\\
48	0.162175\\
49	0.2363375\\
50	0.27606\\
51	0.2853725\\
52	0.291765\\
53	0.29201\\
54	0.29688\\
55	0.3222575\\
56	0.3432875\\
57	0.349185\\
58	0.354335\\
59	0.3677575\\
60	0.383105\\
61	0.3943975\\
62	0.398995\\
63	0.4216075\\
64	0.4513875\\
65	0.4663075\\
66	0.473305\\
67	0.481235\\
68	0.4832575\\
69	0.4875525\\
70	0.4896125\\
71	0.4897375\\
72	0.4898775\\
73	0.4906325\\
74	0.49127\\
75	0.4942425\\
76	0.49464\\
77	0.4946925\\
78	0.4949\\
79	0.4969525\\
80	0.49715\\
81	0.49745\\
82	0.4987225\\
83	0.49965\\
84	0.49979\\
85	0.499895\\
86	0.49991\\
87	0.49992\\
88	0.4999275\\
89	0.499945\\
90	0.4999575\\
91	0.49996\\
92	0.49998\\
93	0.4999825\\
94	0.4999925\\
95	0.4999925\\
96	0.4999925\\
97	0.499995\\
98	0.4999975\\
99	0.4999975\\
100	0.4999975\\
101	0.4999975\\
102	0.4999975\\
103	0.5\\
104	0.5\\
105	0.5\\
106	0.5\\
107	0.5\\
108	0.5\\
109	0.5\\
110	0.5\\
111	0.5\\
112	0.5\\
113	0.5\\
114	0.5\\
115	0.5\\
116	0.5\\
117	0.5\\
118	0.5\\
119	0.5\\
120	0.5\\
121	0.5\\
122	0.5\\
123	0.5\\
124	0.5\\
125	0.5\\
126	0.5\\
127	0.5\\
128	0.5\\
};
\addlegendentry{MK-Polar Codes}

\addplot [color=green, mark=+, mark options={solid, green}]
  table[row sep=crcr]{%
1	0\\
2	0\\
3	0\\
4	0\\
5	0\\
6	0\\
7	0\\
8	0\\
9	0\\
10	0\\
11	0\\
12	0\\
13	0\\
14	0\\
15	0\\
16	0\\
17	0\\
18	0\\
19	0\\
20	0\\
21	0\\
22	0\\
23	0\\
24	0\\
25	0\\
26	0\\
27	0\\
28	2.5e-06\\
29	7.5e-06\\
30	7.5e-06\\
31	2.75e-05\\
32	5.5e-05\\
33	0.0001075\\
34	0.00024\\
35	0.000475\\
36	0.00094\\
37	0.001655\\
38	0.00293\\
39	0.00502\\
40	0.0081525\\
41	0.0128025\\
42	0.01936\\
43	0.0286325\\
44	0.0408875\\
45	0.0574325\\
46	0.0775875\\
47	0.1010475\\
48	0.15095\\
49	0.179535\\
50	0.183525\\
51	0.2184225\\
52	0.2555675\\
53	0.29175\\
54	0.32584\\
55	0.3582225\\
56	0.387685\\
57	0.41313\\
58	0.4346\\
59	0.4517875\\
60	0.466015\\
61	0.4753775\\
62	0.4832\\
63	0.4890275\\
64	0.493185\\
65	0.4949\\
66	0.4953\\
67	0.4971075\\
68	0.4982725\\
69	0.49901\\
70	0.499485\\
71	0.49973\\
72	0.4998475\\
73	0.49991\\
74	0.4999575\\
75	0.499985\\
76	0.4999875\\
77	0.4999925\\
78	0.499995\\
79	0.4999975\\
80	0.4999975\\
81	0.4999975\\
82	0.4999975\\
83	0.4999975\\
84	0.4999975\\
85	0.4999975\\
86	0.5\\
87	0.5\\
88	0.5\\
89	0.5\\
90	0.5\\
91	0.5\\
92	0.5\\
93	0.5\\
94	0.5\\
95	0.5\\
96	0.5\\
97	0.5\\
98	0.5\\
99	0.5\\
100	0.5\\
101	0.5\\
102	0.5\\
103	0.5\\
104	0.5\\
105	0.5\\
106	0.5\\
107	0.5\\
108	0.5\\
109	0.5\\
110	0.5\\
111	0.5\\
112	0.5\\
113	0.5\\
114	0.5\\
115	0.5\\
116	0.5\\
117	0.5\\
118	0.5\\
119	0.5\\
120	0.5\\
121	0.5\\
122	0.5\\
123	0.5\\
124	0.5\\
125	0.5\\
126	0.5\\
127	0.5\\
128	0.5\\
};
\addlegendentry{MK-PAC Codes}

\end{axis}

\end{tikzpicture}%
  \caption{The average TVD of each bit-channel for BE-WTC with $p=0.4$ and $n=128$.}
  \label{fig:TVD}
\end{figure}
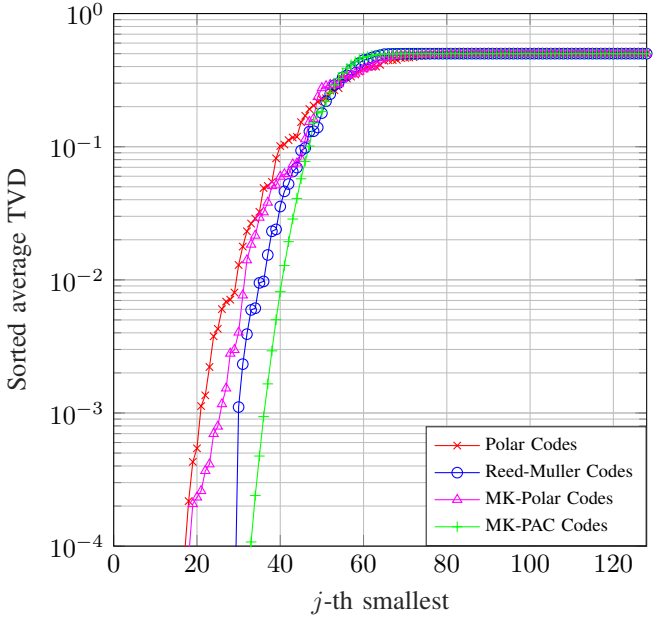


In Fig.~\ref{fig:TVD}, the average TVD of each bit-channel for MK-PAC codes are presented, along with those of polar, Reed--Muller, and MK-polar codes, where the bit-channel indices are sorted with respect to the average TVDs of the bit-channels.
For the MK-PAC codes, the generator polynomial of the convolutional code and the generator matrix of the inner MK-polar code are selected as $p(\const{D})=1+\const{D}^3+\const{D}^7+\const{D}^9+\const{D}^{11}+\const{D}^{12}$ and $\mat{G}_{\textnormal{MK-polar}} = \mat{G}_8 \otimes \mat{G}_{16}$, respectively, where
\begin{IEEEeqnarray*}{c}
    \mat{G}_8=
  \begin{bmatrix}
    \begin{array}{c}
      1 0 0 0 0 0 0 0
      \\ 
      1 1 0 0 0 0 0 0
      \\
      1 0 1 0 0 0 0 0
      \\
      1 0 0 1 0 0 0 0
      \\
      1 1 1 0 1 0 0 0
      \\
      1 1 0 1 0 1 0 0
      \\
      1 0 1 1 0 0 1 0
      \\
      1 1 1 1 1 1 1 1
      \\ 
    \end{array}
  \end{bmatrix}\textnormal{ and }
  \mat{G}_{16}=
  \begin{bmatrix}
    \begin{array}{c}
      0 0 0 0 0 0 0 0 0 0 0 0 0 0 0 1 \\
      0 0 0 0 0 0 0 1 0 0 0 0 0 0 0 1 \\
      0 0 0 0 0 0 0 0 0 0 0 1 0 0 0 1 \\
      0 0 0 0 0 0 0 0 0 0 0 0 0 1 0 1 \\
      0 0 0 0 0 0 0 0 0 0 0 0 0 0 1 1 \\
      0 0 0 0 0 0 0 0 0 0 1 1 0 0 1 1 \\
      0 0 0 0 0 0 0 0 0 0 0 0 1 1 1 1 \\
      0 0 0 1 0 0 0 1 0 0 0 1 1 1 1 0 \\
      0 0 0 0 0 0 1 1 0 0 0 0 0 0 1 1 \\
      0 0 0 0 0 0 1 1 0 1 1 0 0 1 0 1 \\
      0 0 0 0 0 1 0 1 0 0 1 1 1 0 0 1 \\
      0 1 0 1 0 1 0 1 0 1 0 1 0 1 0 1 \\
      0 0 1 1 0 0 1 1 0 0 1 1 0 0 1 1 \\
      0 0 0 0 1 1 1 1 0 0 0 0 1 1 1 1 \\
      0 0 0 0 0 0 0 0 1 1 1 1 1 1 1 1 \\
      1 1 1 1 1 1 1 1 1 1 1 1 1 1 1 1 \\
    \end{array}
  \end{bmatrix}
\end{IEEEeqnarray*}
are taken from \cite{FazeliVardy14_1} (i.e., binary polarization kernels $\mat{K}_8$ and $\mat{K}_{16}$ in \cite{FazeliVardy14_1}). We observe from Fig.~\ref{fig:TVD} that MK-PAC codes have higher polarization speed compared to polar, Reed--Muller, and MK-polar codes.


\subsection{Lower Bounds on the Maximal Secrecy Rate}
\label{sec:secrecy-rates}

To evaluate several achievable secrecy rates, we consider the following upper bounds on the average TVD,  called bound 1 and bound 2, respectively~\cite{Shakiba-HerfehLuzziChorti21_1}:
\begin{IEEEeqnarray}{rCl}
  \IEEEeqnarraymulticol{3}{l}{%
    \bigdTV{P_{M,{\vect{Z}}}}{\mathrm{U}_{\set{M}} P_{\vect{Z}}}
  }\nonumber\\*\quad%
  & \leq &\sum_{j=1}^{k} \BigdTV{P_{\vect{U}_{\{ i_1, \ldots, i_j \}},{\vect{Z}}}}{\frac{1}{2}P_{\vect{U}_{\{ i_1, \ldots, i_{j-1} \}},{\vect{Z}}}},
  \label{eq:TVDbound1}  
\end{IEEEeqnarray}
and
\begin{IEEEeqnarray}{c}
  \bigdTV{P_{M,{\vect{Z}}}}{\mathrm{U}_{\set{M}} P_{\vect{Z}}}\leq\sum_{j=1}^{k} \bigdTV{P_{\vect{U}_{[1:i_j]},{\vect{Z}}}}{\frac{1}{2}P_{\vect{U}_{[1:i_j-1]},{\vect{Z}}}},
  \IEEEeqnarraynumspace\label{eq:TVDbound2}
\end{IEEEeqnarray}
where $i_1 < i_2 < \cdots < i_k$ are the positions of the message bits, i.e., $\set{A} = \{ i_1, i_2, \ldots, i_k \}$. As noted in Section~\ref{sec:avgTVD}, the bit-channels, by abuse of notation, either $\mat{W}^{(i_j)}(\vect{z}, \vect{u}_{\{ i_1, \ldots, i_{j-1} \}}\mid u_{i_j})$ or $\mat{W}^{(i_j)}(\vect{z}, \vect{u}_{[1:i_j-1]}\mid u_{i_j})$, $j \in [1:k]$, are BECs when $\mat{W}$ is a BEC. Thus, we can write the bounds (\ref{eq:TVDbound1}) and (\ref{eq:TVDbound2}) as follows:
\begin{equation}
  \bigdTV{P_{M,\vect{Z}}}{\mathrm{U}_{\set{M}} P_{\vect{Z}}}\leq\frac{1}{2} \sum_{j=1}^{k} (1 - \tilde{p}_j), \label{eq:TVDbound-BE-WTC}
\end{equation}
where $\tilde{p}_j$ is the erasure probability of the bit-channel $\mat{W}^{(i_j)}(\vect{z}, \vect{u}_{\{ i_1, \ldots, i_{j-1} \}}\mid u_{i_j})$ or $\mat{W}^{(i_j)}(\vect{z},\vect{u}_{[1:i_j-1]}\mid u_{i_j})$. Given a value of the secrecy leakage constraint $\delta$ on the right-hand side of \eqref{eq:TVDbound-BE-WTC}, we can determine the maximum number of bit-channels, denoted by $\tilde{k}$, such that their total sum of average TVDs is not greater than $\delta$, i.e., $\edTV{P_{M,\vect{Z}}}{\mathrm{U}_{\set{M}} P_{\vect{Z}}} \leq\frac{1}{2}\sum_{j=1}^{k}(1-\tilde{p}_j)\leq\delta$. This leads to a lower bound $\tilde{k}/n$ on the maximal secrecy rate $\const{R}^\ast(n,\delta)$ since by Definition~\ref{def:def_maximal-secrecy-rate}, we have
\begin{IEEEeqnarray*}{rCl}
  \IEEEeqnarraymulticol{3}{l}{%
    \const{R}^\ast(n,\delta)}\nonumber\\*%
  & \geq &\max\biggl\{\frac{k}{n}:\exists\,(n,2^{k}) \textnormal{ code such that }\frac{1}{2}\sum_{j=1}^{k}(1-\tilde{p}_j)\leq\delta\biggr\}.
\end{IEEEeqnarray*}

\begin{figure}[t!]
  \centering
  \input{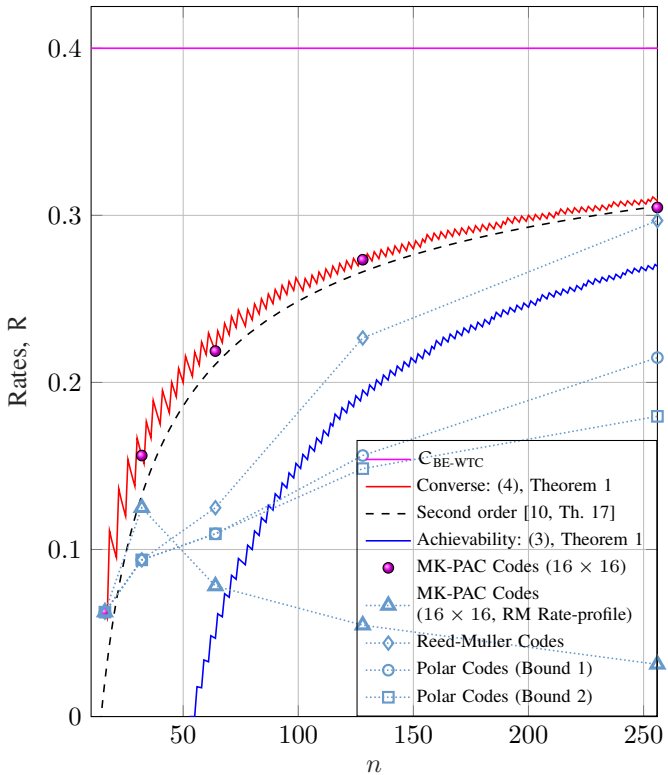}
  \caption{Code performance on semi-deterministic BE-WTC with $p=0.4$ and $\delta=0.001$.}
  \label{fig:secrecy-rates_delta1e-3}
\end{figure}
\begin{figure}[t!]
  \centering
  \input{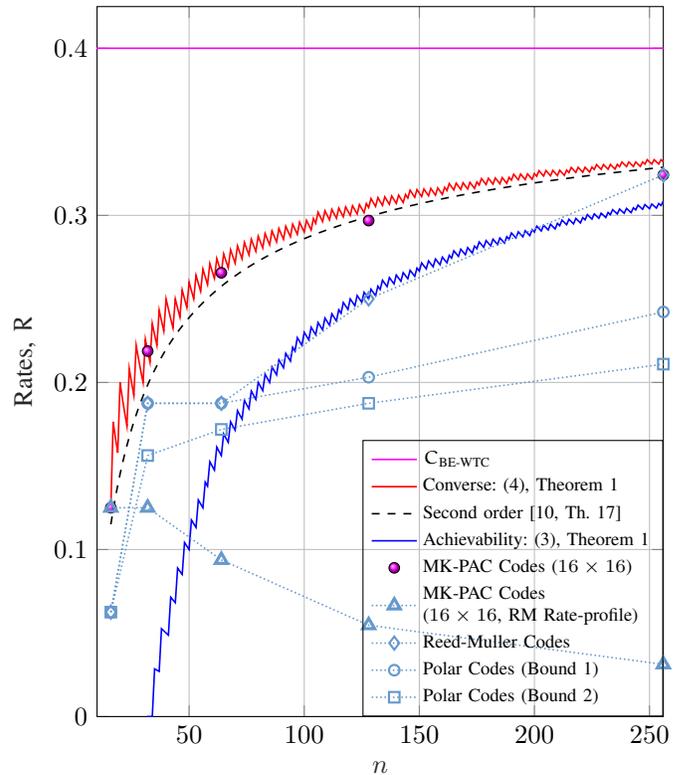}
  \caption{Code performance on semi-deterministic BE-WTC with $p=0.4$ and $\delta=0.01$.}
  \label{fig:secrecy-rates_delta1e-2}
\end{figure}

In Fig.~\ref{fig:secrecy-rates_delta1e-3}, the lower bounds on the maximal secrecy rate obtained from polar, Reed--Muller, and MK-PAC codes are presented for the case of $\delta = 0.001$ on BE-WTC, along with the second order approximation secrecy rate, the random coding achievability (\eqref{eq:achievability_BEWTC}, Theorem~\ref{thm:BE-WTC-nonasymptotics}), and the exact converse bound (\eqref{eq:converse-bound_BEWTC}, Theorem~\ref{thm:BE-WTC-nonasymptotics}). For comparison, the results for MK-PAC codes with the Reed-Muller (RM) rate-profile~\cite{Arikan19_1sub} are also shown. The zigzagging behavior of the plot is common to the achievability and converse bounds as in the simulation, we make $\log_2{\const{M}}=k$ be integer values. For MK-PAC codes, the generator matrices are selected as $\mat{G}_{16}$, $\mat{G}_2 \otimes \mat{G}_{16}$, $\mat{G}_2 \otimes \mat{G}_2 \otimes \mat{G}_{16}$, $\mat{G}_8 \otimes \mat{G}_{16}$, and $\mat{G}_{16} \otimes \mat{G}_{16}$ for $n = 16, 32, 64, 128$, and $256$, respectively. For $n=16$ and $32$, the outer convolutional codes have $p(\const{D})=1+\const{D}^2+\const{D}^3+\const{D}^5+\const{D}^6$, while for $n=64, 128$, and $256$, the generator polynomials are selected as $p(\const{D})=1+\const{D}^3+\const{D}^7+\const{D}^9+\const{D}^{10}$, $p(\const{D})=1+\const{D}^3+\const{D}^7+\const{D}^9+\const{D}^{11}+\const{D}^{12}$, and $p(\const{D})=1+\const{D}+\const{D}^3+\const{D}^6+\const{D}^{10}+\const{D}^{12}+\const{D}^{15}+\const{D}^{17}+\const{D}^{18}$, respectively. We observe from Fig.~\ref{fig:secrecy-rates_delta1e-3} that MK-PAC codes show promising performance in the BE-WTC, specifically for blocklengths 16, 32, 64, 128, and 256. In particular, under the average TVD secrecy metric, MK-PAC codes can achieve secrecy rates beyond the second-order approximation rate for short blocklengths. More remarkably, we observe that MK-PAC codes can achieve the optimal secrecy rate, i.e., the converse bounds for the maximal secrecy rate, at blocklengths $n=16, 32, 64$, and $128$, exactly match the achievable secrecy rates of MK-PAC codes. Observations similar to the above can also be made in Fig.~\ref{fig:secrecy-rates_delta1e-2}, which depicts the lower bounds on the maximal secrecy rate for the case of $\delta = 0.01$ on BE-WTC.

\section{Conclusion}
\label{sec:conclusion}
In this paper, we consider the semi-deterministic binary erasure wiretap channel, where the main channel to the legitimate receiver is noiseless, and the eavesdropper’s channel is a BEC, and investigate the problem of achieving the optimal short-blocklength secrecy rate. We consider the average TVD as the secrecy metric and derive the non-asymptotic theoretical bounds on the maximal secrecy rate. We further provide the achievable secrecy rates of MK-PAC codes and compare them to the second-order secrecy rates, the random coding achievability, and the exact converse bounds. Numerical results indicate that under the average TVD secrecy metric, MK-PAC codes can achieve secrecy rates beyond the second-order approximation rate for short blocklengths. More notably, we observe that MK-PAC codes can also achieve the maximal secrecy rate at certain blocklengths when the secrecy leakage does not exceed several values.



\IEEEtriggeratref{13}

\bibliographystyle{IEEEtran}
\bibliography{defshort1,biblioHY} 









\end{document}